%% file: main.tex
\documentclass[sigconf]{aamas}

\usepackage{algorithmic}
\usepackage{graphicx}
\usepackage{textcomp}
\usepackage{xcolor}
\usepackage{hyperref}
\usepackage{tikz}
\usepackage{relsize}
\usepackage{enumitem}
\usepackage[commandnameprefix=ifneeded]{changes}

\makeatletter
\gdef\@copyrightpermission{
  \begin{minipage}{0.2\columnwidth}
   \href{https://creativecommons.org/licenses/by/4.0/}{\includegraphics[width=0.90\textwidth]{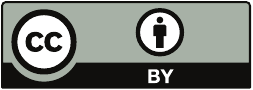}}
  \end{minipage}\hfill
  \begin{minipage}{0.8\columnwidth}
   \href{https://creativecommons.org/licenses/by/4.0/}{This work is licensed under a Creative Commons Attribution International 4.0 License.}
  \end{minipage}
  \vspace{5pt}
}
\makeatother

\setcopyright{ifaamas}
\acmConference[AAMAS '25]{Proc.\@ of the 24th International Conference
on Autonomous Agents and Multiagent Systems (AAMAS 2025)}{May 19 -- 23, 2025}
{Detroit, Michigan, USA}{Y.~Vorobeychik, S.~Das, A.~Nowé  (eds.)}
\copyrightyear{2025}
\acmYear{2025}
\acmDOI{}
\acmPrice{}
\acmISBN{}

\acmSubmissionID{1026}

\title{Uncertainty Expression for Human-Robot Task Communication}

\author{David Porfirio}
\affiliation{
  \institution{U.S. Naval Research Laboratory}
  \city{Washington, D.C.}
  \country{USA}}
\email{david.j.porfirio2.civ@us.navy.mil}

\author{Mark Roberts}
\affiliation{
  \institution{U.S. Naval Research Laboratory}
  \city{Washington, D.C.}
  \country{USA}}
\email{mark.c.roberts20.civ@us.navy.mil}

\author{Laura M. Hiatt}
\affiliation{
  \institution{U.S. Naval Research Laboratory}
  \city{Washington, D.C.}
  \country{USA}}
\email{laura.m.hiatt.civ@us.navy.mil}

\begin{abstract}

An underlying assumption of many existing approaches to human-robot task communication is that the robot possesses a sufficient amount of environmental domain knowledge, including the locations of task-critical objects. This assumption is unrealistic if the locations of known objects change or have not yet been discovered by the robot. In this work, our key insight is that in many scenarios, robot end users possess more scene insight than the robot and need ways to express it. Presently, there is a lack of research on how solutions for collecting end-user scene insight should be designed. We thereby created an \textit{Uncertainty Expression System} (UES) to investigate how best to elicit end-user scene insight. The UES allows end users to convey their knowledge of object uncertainty using either: (1) a precision interface that allows meticulous expression of scene insight; (2) a painting interface by which users create a heat map of possible object locations; and (3) a ranking interface by which end users express object locations via an ordered list. We then conducted a user study to compare the effectiveness of these approaches based on the accuracy of scene insight conveyed to the robot, the efficiency at which end users are able to express this scene insight, and both usability and task load. Results indicate that the \rank{} interface is more user friendly and efficient than the \precision{} interface, and that the \paint{} interface is the least accurate.

\end{abstract}


\DeclareMathOperator*{\argmin}{argmin}
\newcommand{\BibTeX}{\rm B\kern-.05em{\sc i\kern-.025em b}\kern-.08em\TeX}
\newcommand{\tool}{{\textit{UES}}}
\newcommand{\precision}{{\textit{precision}}}
\newcommand{\paint}{{\textit{paint}}}
\newcommand{\rank}{{\textit{rank}}}
\newcommand{\Precision}{{\textit{Precision}}}
\newcommand{\Paint}{{\textit{Paint}}}
\newcommand{\Rank}{{\textit{Rank}}}

\newcommand{\eg}[0]{\textit{e.g.,}}
\newcommand{\ie}[0]{\textit{i.e.,}}
\newcommand{\hone}[0]{\textbf{H1}}
\newcommand{\htwo}[0]{\textbf{H2}}
\newcommand{\hthr}[0]{\textbf{H3}}
\newcommand{\hfou}[0]{\textbf{H4}}
\newcommand{\hfiv}[0]{\textbf{H5}}

\definecolor{blue}{RGB}{0, 0, 255} 
\newcommand{\bluesquare}[1]{%
    \begin{tikzpicture}[baseline=(char.base)]
        \node[draw=blue, fill=blue, rectangle, rounded corners=2pt, minimum height=10pt, minimum width=10pt, inner sep=0pt, text=white] (char) {\textbf{#1}};
    \end{tikzpicture}%
}
\definecolor{purple}{RGB}{122, 0, 255} 
\newcommand{\purplesquare}[1]{%
    \begin{tikzpicture}[baseline=(char.base)]
        \node[draw=purple, fill=purple, rectangle, rounded corners=2pt, minimum height=10pt, minimum width=10pt, inner sep=0pt, text=white] (char) {\textbf{#1}};
    \end{tikzpicture}%
}
\definecolor{yellow}{RGB}{255, 200, 0} 
\newcommand{\yellowsquare}[1]{%
    \begin{tikzpicture}[baseline=(char.base)]
        \node[draw=yellow, fill=yellow, rectangle, rounded corners=2pt, minimum height=10pt, minimum width=10pt, inner sep=0pt, text=black] (char) {\textbf{#1}};
    \end{tikzpicture}%
}
\definecolor{black}{RGB}{0, 0, 0} 
\newcommand{\whitesquare}[1]{%
    \begin{tikzpicture}[baseline=(char.base)]
        \node[draw=black, fill=white, rectangle, rounded corners=2pt, minimum height=10pt, minimum width=10pt, inner sep=0pt, text=black] (char) {\textbf{#1}};
    \end{tikzpicture}%
}

\begin{document}

\pagestyle{fancy}
\fancyhead{}

\maketitle

\input{1introduction}

\input{2related_work}

\input{3system_design}

\input{4evaluation}

\input{5discussion}

\section{Conclusion}
We investigated three different approaches for communicating uncertain object locations to a robot via a user interface.
We compared each approach based on user experience, how efficient end users can specify uncertain object locations, and the accuracy of the user-specified locations compared to ground truth.
Furthermore, we demonstrated how user-specified object locations can be used to generate task plans.
Our results indicated that a ranking approach to uncertainty expression is both easier and faster for users.

\section*{Acknowledgment}
This research was supported by the U.S. Naval Research Laboratory and an NRC Postdoctoral Research Associateship awarded to DP. The views and conclusions contained herein are those of the authors and should not be interpreted as necessarily representing the official policies, either expressed or implied, of the U.S. Navy.

\bibliographystyle{ACM-Reference-Format} 
\bibliography{ref}

\end{document}

%% file: 1introduction.tex
\section{Introduction}
\begin{figure}[!t]
\centering
    \includegraphics[width=\columnwidth]{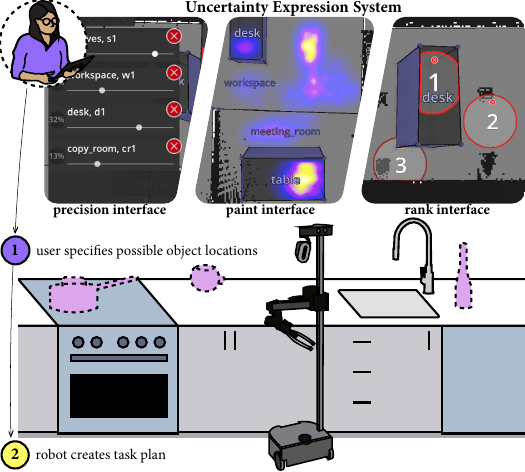}
    \caption{With the \textit{Uncertainty Expression System}, we investigate three different interfaces that enable humans to express possible object locations to the robot. From this scene insight, the robot can create a task plan to achieve an objective.}
    \label{fig:teaser}
\end{figure}

In many human-robot teaming scenarios, the robot is expected to independently complete its task objectives autonomously \textit{at runtime} and without human intervention \cite{harris2021auv}, meaning that critical human domain knowledge should be communicated to the robot \textit{prior} to runtime.
It is crucial that the human has proper tools for conveying this domain knowledge to their robotic counterparts easily, efficiently, and accurately.
Consider, for example, a disaster relief search-and-rescue scenario, in which mobile robots and unmanned aerial vehicles (UAVs) traverse buildings and terrain that are unsafe or unreachable by their human teammates.
If human intervention at runtime is infeasible, all task details must be conveyed prior to runtime.
In this case, a member of the disaster relief team uses an off-the-shelf command interface to specify that the robot should \textit{``search for victims and provide aid.''}

In a fully observable environment, the robot would have a sufficient amount of information to create a plan to autonomously achieve this objective.
In uncertain environments, however, an objective alone may not be enough information for the robot to complete its task efficiently.
Although the robot can recognize novel entities and construct a semantic map of its immediate surroundings, it lacks an initial belief state about where task-critical entities (\eg{} disaster victims) \textit{might} be.

There are several different ways in which the robot can obtain an initial belief state.
In one approach, the robot can uncover scene knowledge by itself though exploring and observing its environment.
Exploration, however, is potentially inefficient without guidance.
System developers, namely the creators of the robot platform, are a potential \textit{a priori} source of this guidance and can provide the robot with \textit{strategies} for finding contextual scene information.
However, these developers are not a good source of deep, up-to-date, contextual knowledge.
Recent work in human-robot tasking instead positions end users as rich sources of this knowledge \cite{stegner2022care}.
We posit that in scenarios where users need autonomous robot assistance but object locations in the environment are unknown, the user may be best situated to provide the robot with this knowledge and needs a way to express it.

We assume that our disaster-relief scenario has unfolded such that the humans have more \textit{scene insight} than the robot for where task-critical entities reside.
If the human can convey this knowledge to the disaster relief robot, the robot might perform its search more efficiently.
Unfortunately, existing approaches to human-robot task communication are limited in enabling humans to convey this knowledge to robots.
Instead, these approaches largely focus on conveying task objectives alone, such as through natural language commands or simple end-user programming scripts, without allowing robot end users to express uncertain entity locations.

We thereby created an \textit{Uncertainty Expression System} (\tool{}) in order to investigate how best to design task communication tools that enable end users to convey uncertain scene insight to the robot.
Shown in Figure \ref{fig:teaser}, the \tool{} contains three separate approaches for enabling end users to specify the possible locations of task-critical objects---a (1) \precision{} interface via which end users convey scene insight through the exact probability distributions of where objects are believed to be; a (2) \paint{} interface via which end users convey these distributions by painting a heat map; and (3) a \rank{} interface through which users specify lists of possible locations ordered from most to least likely to contain task-critical objects.

In creating and investigating three different approaches for expressing scene insight, our primary research aim is exploratory.
Specifically, we aim to investigate the advantages and disadvantages of each approach in order to inform how \textit{uncertainty expression tools}, namely interfaces that enable end users to convey scene insight to a robot, should be designed.
Our research questions are, \textit{how should uncertainty expression tools be designed to: (RQ1) maintain accuracy of user-specified probability distributions of object locations in the robot's environment; (RQ2) increase the efficiency at which these distributions are conveyed to the robot; and (RQ3) maintain user experience in terms of tool usability and cognitive load?}

Our contributions are as follows:
\begin{itemize}
    \item \textit{System}---the \textit{Uncertainty Expression System}, including \precision{}, \paint{}, and \rank{} interfaces for conveying scene insight to a robot.
    \item \textit{Empirical}---a user study that assesses the effectiveness of each interface in terms of accuracy of user-specified uncertain object locations, efficiency at which these locations can be specified, user experience, and cognitive load. We additionally include a case study that investigates how scene insight provided from each interface might impact robot performance at runtime. 
    \item \textit{Design}---several implications that pertain to the integration of human input into robotic systems and best practices to allow users to interact with these systems. We find that \rank{} outperforms \precision{} and \paint{} in user experience. Moreover, although \paint{} is less accurate, a case study indicates that it does not necessarily pose a risk to task performance. 
\end{itemize}

%% file: 2related_work.tex
\section{Related Work}

Our work relates to prior research in human-robot task communication, robot scene understanding, and uncertainty elicitation.

\subsection{Human-Robot Task Communication}
This work pertains to scenarios in which a human communicates a necessary and sufficient amount of task detail to a robot \textit{before} runtime, after which the robot autonomously completes the task \textit{at} runtime (in contrast to shared autonomy or teleoperation) \cite{sebo2024autonomy}.
Such scenarios usually involve \textit{command} interfaces (\eg{} via natural language as in \cite{Matuszek2013}) or \textit{end-user programming} interfaces \citep[\eg{}][]{porfirio2024goal} through which the user communicates the objectives of a task via a set of actions or goals for the robot to complete. Given a set of actions or goals, the robot can plan its task accordingly via a combination of task and motion planning \cite{ghallab2016planning,kaeb2021tamp}, control methods \citep[\eg{}][]{wong2017control}, and executing individual skills using learned policies \citep[\eg{}][]{patra2022curriculum}. In uncertain environments, the robot may employ probabilistic planning approaches such as \textit{FF-Replan} \cite{yoon2007ff} or \textit{POMCP} \cite{silver2010monte}.

Most human-robot task communication research focuses on enabling humans to express task \textit{objectives} rather than supporting domain information, including any \textit{uncertainty} in the robot's environment.
Some approaches leverage object locations as concrete referents \cite{sharma2022correcting} or enable humans to tell the robot where an object is with certainty (\eg{} ``grab the groceries \textbf{\textit{from the table}}'') \cite{walker2019semantic}. Others pair natural language with other input modalities, such as gesture, to resolve ambiguity and further convey exactness \cite{weerakoon2020gesture}.
Many interfaces either assume that the robot has visual access to task-critical objects (as is often the case for closed collaborative environments \cite{schoen2022coframe,huang2017code3,ulas2024alchemis}), has previously encountered them, or is at least capable of finding them  \cite{porfirio2024goal,porfirio2023sketch,huang2016custom,huang2020vipo}.
Users are thereby unable to convey a belief about where objects \textit{might} be.

\subsection{Scene Understanding in Robotics}

Semantic scene understanding concerns the ability of a robot to analyze its environment and understand the relationships of entities within \cite{patel2023scene}.
Recent work in object recognition (\eg{} \textit{YOLO} \cite{redmon2016yolo}) and semantic mapping of places \citep[\eg{}][]{sunderhauf2016mapping} have enabled robots to recognize and interact with novel objects in a zero-shot manner \cite{ha2022semantic,liu2024okrobot}. As a result, if the robot can view and explore its environment, it can interact with it.

In practice, the robot may create or be given a full semantic representation of its environment prior to being asked to complete a task \cite{bolte2023usanet}.
Even so, the locations of objects in these environments may change frequently, or the robot may be unable to generate an accurate representation of its environment prior to being assigned a task.
The robot must then find task-critical objects dynamically by simultaneously performing its task while under partial uncertainty.
Prior approaches have incorporated uncertainty within the task and motion planner's cost function \cite{gualtieri2021pickplace} or configured a \textit{belief space} that the robot updates as it executes its task and receives new information \cite{kaebling2013belief,silver2010monte}.
In the latter case, the robot may still need or benefit from being provided an initial belief state.
There is a lack of research in best approaches for enabling humans to convey this belief state as \textit{scene insight} to their robotic counterparts. 

\subsection{Uncertainty Elicitation}

Our key insight is that humans are often well-positioned to express beliefs to the robot.
Belief expression is a key objective of \textit{probability elicitation}, in which users express their belief about the probabilities of some phenomena occurring \cite{savage1971probability}.
Notable examples of existing elicitation tools include \textit{MATCH}, a web-based interface that provides five straightforward approaches for directly eliciting continuous distributions from experts (\ie{} asking experts to approximate the distribution) \cite{morris2014match}; and Elicitator, an expert landscape ecology tool that uses \textit{indirect} elicitation to learn a probabilistic model from expert knowledge \cite{lowchoy2010elicitator}. Many other existing tools require insight from multiple domain experts and are subject to biases \cite{ohagan2019biases}. Contrary to existing tools, we seek an approach for dynamic belief expression of categorical, rather than continuous, distributions of object locations.

Still, the \tool{} mirrors some common paradigms in both uncertainty elicitation and task specification.
The \precision{} interface is akin to the \textit{probability scale} approach of uncertainty elicitation, a simple method in which experts are asked to specify the exact probability of an event on a linear interval between $0\%$ and $100\%$ \cite{renooij2001elicit}.
The \paint{} and \rank{} interfaces, by contrast, are inspired from existing literature in end-user development and human-robot interaction task specification.
\Paint{} draws from literature on sloppy programming \cite{little2010sloppy}, which prioritizes user experience at the expense of precision, and is additionally inspired from the success of \textit{sketching} interfaces in human-robot interaction \cite{sakamoto2009sketch, liu2011roboshop,porfirio2023sketch}.
\Rank{} differs from both \precision{} and \paint{} in that rather than specifying probability distributions, users specify an ordered list of locations that objects are likely to be from most to least likely. 
If interpreted as the order at which the robot should search for the objects, \rank{} is similar to expressing sequences of robot objectives, a paradigm common to end-user development \cite{porfirio2024goal}.
It is unclear which approach performs best in terms of user experience and robot task performance. 

%% file: 3system_design.tex
\section{System Design}\label{sec:design}

With the \tool{}, robot end users can specify uncertain scene insight to the robot in the form of the possible locations of task-critical objects.
For our purposes, \textit{location} informally refers to some $(x,y)$ position in a top-down representation of the robot's environment.
To receive this knowledge, the robot should be equipped with \textit{preliminary scene information} in the form of a map of the environment in which key regions and surfaces have been labeled (\ie{} automatically labeled by the robot or hand-labeled by the end user).
In our description of the \tool{}, we begin by describing this preliminary scene information in greater depth.
Then we describe each of the three uncertainty expression interfaces offered by the \tool{}---the \precision{} interface, the \paint{} interface, and the \rank{} interface.

Throughout we refer to a delivery example to help illustrate the system.
Consider a mobile delivery robot that exists in a professional workplace.
The robot has the ability to localize itself and travel between different points in the workplace, in addition to being able to recognize objects that it can interact with.
The robot can grab these objects and put them in different locations.

Now consider the following delivery command that a user provides to the robot in natural language: \textit{``Place an umbrella in my co-worker's bag.''}
Suppose that the robot can recognize the umbrella and the bag when it sees them, but both objects are out of its view, and it does not know where it should begin looking.
Suppose that the user, by contrast, has seen these objects in several different locations in the workplace.
Knowing the user's belief of where task-critical objects exist can assist the robot in efficiently completing its task. 
The role of the \tool{} is therefore to help the user express these probable locations to the robot.

\subsection{The Robot's Preliminary Scene Information}

\begin{figure}[!t]
    \centering
    \includegraphics[width=\columnwidth]{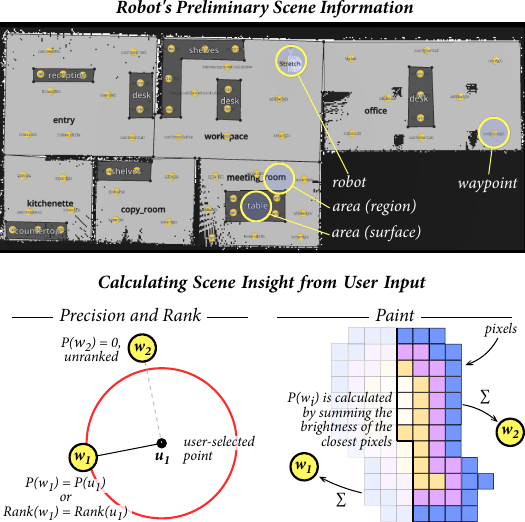}
    \caption{(Top) Preliminary scene information that the robot must possess. (Bottom left) scene insight from the \precision{} and \rank{} interfaces are calculated by determining the closest waypoints to user-selected points. (Bottom right) scene insight from the \paint{} interface is calculated by summing the brightness of pixels closest to each waypoint.}
    \label{fig:map}
\end{figure}

Prior to users conveying object locations to the robot, the robot must be equipped with basic information about the scene.
Specifically, the robot must either create or be provided with a two-dimensional map of its environment via simultaneous localization and mapping (SLAM).
This map is displayed in the \tool{}.
The robot should additionally be equipped with limited semantic information about this scene, including labeled areas that the robot can traverse and places that the robot is able to visit.
In what follows, we describe how the \tool{} represents this preliminary scene information.

\subsubsection{Areas}

We define an \textit{area} as a planar bounding polygon on the SLAM map that describes either a \textit{region} or a \textit{surface}.
Regions are areas that are traversable by the robot, such as rooms and cooridors.
In the delivery example, regions include the \textit{meeting room}, \textit{office}, \textit{copy room}, and other areas designated in light grey in Figure \ref{fig:map} (top).
Conversely, surfaces are areas that are not traversable by the robot, including obstacles like tables and chairs.
Surfaces in the delivery example include the \textit{shelves}, \textit{desks}, the \textit{countertop}, and other areas designated in dark grey in Figure \ref{fig:map} (top).

Task-critical objects can exist on areas.
For ease of quantifying all possible points on an area that an object might exist in, the whole map is discretized into \textit{pixels} on a cartesian coordinate system.
We define $X_a$ as the set of all pixels in area $a \in A$. We then define $X$ as the set of all pixels on the map, $X = \bigcup_{a \in A}^{} X_a$.

\subsubsection{Waypoints}

We define \textit{waypoints} as points in the scene that the robot can visit and travel between.
Waypoints are conceptually similar to prior work \cite{hawes2011homealone, hanheide2017world} as nodes in a topological navigation graph.
Waypoints are invisible to the user but are shown for convenience as yellow dots in Figure \ref{fig:map} (top).
Similar to areas, waypoints must be known to the robot prior to receiving scene insight in the form of uncertain object locations from the human.

Waypoints are additionally used to describe the potential locations at which an object might reside.
If an object resides closest to a particular waypoint, the object is said to be \textit{at that waypoint}.
We thereby define $X_w$ as the set of pixels closest to waypoint $w \in W$.
Formally, the pixels adjacent to $w$ can be expressed as a Voronoi cell, $X_w = \{x \in X \mid w = \argmin_{w_i \in W} ||w_i-x||\}$, where $||w_i - x||$ is the distance between positions $w_i$ and $x$ in two-dimensional space. 

\subsection{Expressing Human scene insight}

\begin{figure*}[!t]
    \centering
    \includegraphics[width=\textwidth]{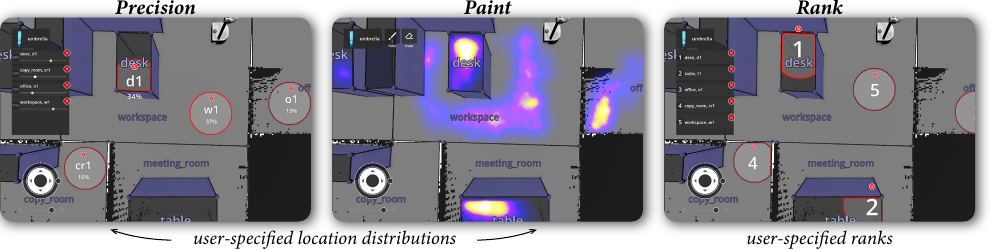}
    \caption{The three \tool{} interfaces. \Precision{} (left): users select points where objects may exist and use sliders to express their probabilities. \Paint{} (center): users express probabilities by painting a heat map. \Rank{} (right): users select and rank points from most to least likely. For \precision{} and \rank{}, points are visualized as red circles. Circles that do not fit on an area are cropped.}
    \label{fig:ui}
\end{figure*}

Equipped with this preliminary scene information, users convey uncertain object locations to the robot via the \tool{}.
The \tool{} displays the SLAM map and labeled areas to the user.
Users can adjust the view of the SLAM map by zooming in and out and panning laterally. 
Scene insight is then conveyed to the robot via one of three interfaces---\precision{}, \paint{}, and \rank{}---which we describe below.

\subsubsection{\Precision{} Interface}\label{sec:design_precision}

The \precision{} interface (Figure \ref{fig:ui}, left) allows users to convey categorical probability distributions of object locations.
To specify a probability distribution, the user first selects points on the map.
Each user-selected point maps to a single pixel and is visually depicted as a red circle with the pixel being its center.
The user then uses slider bars to set how likely the user believes the object to exist at each point.
User-selected points are \textit{free-floating}. That is, points do not ``snap'' to waypoints (\ie{} their locations are not adjusted \textit{post hoc}), and users are unaware of which waypoints map to their selections.
In this way, the \precision{} interface is intended to capture the user's true, unadjusted understanding of the scene, agnostic to the set of waypoints underneath.

Each slider $s_u$ corresponding to user-selected point $u$ exists on the interval $[0, 1]$.
Let $Pr(u)$ be the probability of an object existing at $u$.
If the sum of sliders is greater than $1.0$, $Pr(u)$ is normalized.
Otherwise, $Pr(u)$  is not normalized.
In this way, we allow users to specify incomplete distributions where $\sum_u Pr(u) < 1.0$:
\begin{align*}
Pr(u) = \begin{cases} 
          s_u & \mathrm{if} \sum_{i}^{}s_i <= 1.0,\\
          \frac{s_u}{\sum_{i}^{}s_i} & \mathrm{otherwise.} 
       \end{cases}
\end{align*}

\paragraph{Calculating a probability distribution over waypoints} 
The \precision{} interface produces a categorical distribution of an object existing at different waypoints.
Figure \ref{fig:map} (bottom left) illustrates how this distribution is calculated.
Let $X_u$ be the set of user-selected points, where $u \in X_u$ maps to a single pixel.
Recall that $X_w$ is the set of pixels closest to waypoint $w$ and let $V=X_u\cap X_w$ be the set of user-selected points closest to $w$.
The probability of an object being at $w$ is thus the weighted sum of user-selected points near $w$:
\begin{align*}
    Pr(w) = \mathlarger{\sum}\limits_{v \in V}^{} Pr(v).
\end{align*}

\subsubsection{\Paint{} Interface}

Shown in Figure \ref{fig:ui} (center), the \paint{} interface similarly allows users to convey probability distributions of object locations.
To specify a probability distribution, the user clicks and drags their cursor to paint a heat map.
Heat maps can be drawn on both regions and surfaces.
Clicking for more time at a particular area causes the heat map to grow brighter in that area, with the brightness of each pixel $x \in X$ being a scalar value $b_x \in [0, 1]$.
Brightness values closer to $1$ correspond to higher likelihoods the object is to exist in those areas.

Pixels are assigned colors based on their brightness, interpolated between lower brightness being assigned \bluesquare{blue}, mid-range brightness being assigned \purplesquare{violet}, higher brightness being assigned \yellowsquare{yellow}, and the brightest values being assigned \whitesquare{white}. The \bluesquare{blue}-\purplesquare{violet}-\yellowsquare{yellow} color palette was chosen due to being distinguishable for individuals with and without color vision deficiencies.

\paragraph{Calculating a probability distribution over waypoints} Similar to the \precision{} interface, user input to the \paint{} interface results in a categorical probability distribution of an object existing at different waypoints on the map. Recall that $X_{w}$ is the set of pixels closest to waypoint $w \in W$. The probability of an object existing at $w$ is therefore the summed brightness of \textit{adjacent} pixels divided by the summed brightness of \textit{all} pixels (Figure \ref{fig:map}, bottom right):
\begin{align*}
\Pr(w) = \frac{
    \sum\limits_{x \in X_{w}}^{} b_{x}
}{
    \sum\limits_{y \in X}^{} b_y
}.
\end{align*}

\subsubsection{\Rank{} Interface}

Shown in Figure \ref{fig:ui} (right), the \rank{} interface is similar to the \precision{} interface in that users select points on the environment, and these points are similarly free-floating in order to represent users' true, unadjusted understanding of a scene.
In contrast, however, users do not specify a probability distribution of object locations.
Rather, they specify the order of locations at which they believe the object to exist, with the highest rank corresponding to the object's most likely location, and the lowest rank corresponding to the object's least likely location out of the points specified by the user.
Users can rearrange ranks in the interface.

%% file: 4evaluation.tex
\section{Evaluation}

We conducted an IRB-approved human-subjects study to compare the accuracy, efficiency, and user experience of each uncertainty expression interface offered by the \tool{}.
Our evaluation is intended to be exploratory, and our primary goal is to uncover the advantages and disadvantages of each interface.
Our hypotheses are as follows. 
\begin{itemize}[leftmargin=12pt]
    \item[] \hone{}: The choice of interface will affect accuracy, efficiency, user experience, and task load.
    \item[] \htwo{}: Usability will be highest with the \paint{} interface and lowest with the \precision{} interface. This hypothesis pertains to RQ3 and is informed by the \paint{} interface having the simplest visuals and controls compared to the \precision{} interface.
    \item[] \hthr{}: Cognitive load will be lowest with the \paint{} interface and highest with the \precision{} interface. This hypothesis also pertains to RQ3 and is also informed by the \paint{} and \precision{} interfaces having simple versus complex controls, respectively.
    \item[] \hfou{}: Efficiency will be highest with the \rank{} interface and lowest with the \precision{} interface. This hypothesis pertains to RQ2 and is informed by the keystroke-level model of human-computer interaction \cite{card1980keystroke}, in which \rank{} performs best due to potentially involving the least interaction with the UI.
    \item[] \hfiv{}\textbf{.1}: The \precision{} interface will exhibit higher accuracy than the \paint{} interface when comparing scene insight to ground-truth distributions. \hfiv{}\textbf{.2} In a rank-based comparison of all three interfaces, the \precision{} interface will exhibit the highest accuracy and the \rank{} interface will exhibit the lowest accuracy. These hypotheses pertain to RQ1 and are informed by the \precision{} interface affording users the highest control over exact probability values, whereas the \rank{} interface afford users with the lowest control over these values.  
\end{itemize}

In addition to our main evaluation, we also provide a case study that demonstrates how user-specified scene insight can assist the robot in planning to achieve its objective.
Our case study evaluates whether robot performance might be affected by the choice of interface.
As the case study is mainly intended for demonstration, we do not include any hypotheses pertaining to robot performance.

\subsection{Procedure}\label{sec:procedure}

\begin{figure*}[!t]
    \centering
    \includegraphics[width=\textwidth]{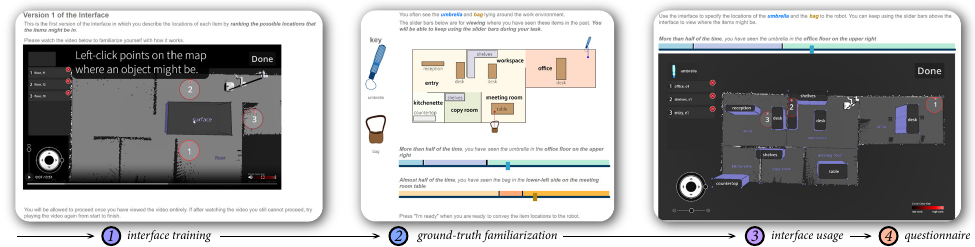}
    \caption{The procedure for each trial involved four steps. First, (1) participants watched a tutorial video on a particular interface. Next, (2) participants became familiarized with ground-truth probabilities of objects being in different locations. Then, (3) participants used the interface to express scene insight to the robot. Lastly, (4) participants answered a questionnaire.}
    \label{fig:procedure}
\end{figure*}

The study procedure was conducted online via Qualtrics\footnote{\url{https://www.qualtrics.com/}} and distributed to members of our research organization.
Our evaluation was within-subjects, in which every participant in the study used all three of the interfaces.
Individuals who clicked on the link were first screened for eligibility and device compatibility.
Individuals were then provided with an informed consent document.

Upon providing consent, participants were familiarized with the office scenario described in section \S\ref{sec:design}, a hypothetical robot (the Hello Robot \textit{Stretch} \cite{kemp2022stretch}), and the task for the robot to complete.
Specifically, participants were briefed on the task described in section \S\ref{sec:design}: the robot must place an umbrella in someone's bag, but the robot is unaware of where the umbrella and the bag exist.
Participants were informed that they must convey to the robot the likely locations of these items.
Following the task briefing, participants were exposed to a short tutorial about the \tool{}'s basic controls.

After the tutorial, participants underwent three trials to address the scenario, each trial using a different interface.
The interface order followed a complete counterbalancing scheme, falling two participants short of full counterbalancing.
Figure \ref{fig:procedure} illustrates the four steps involved in each trial: (1) training on interface usage; (2) ground-truth familiarization; (3) interface usage; and (4) questionnaire response.
Training involved watching a 1-minute tutorial video about the interface.
Ground truth familiarization involved interacting with a visual mockup of the office environment with three possible locations each for the umbrella and bag, including how likely these items might be in each location.
The visual mockup was accompanied by natural language descriptions of where each object might be and how likely it would be there.
To prevent the need for memorizing ground truth locations and likelihoods, the natural language descriptions and likelihoods for each location remained accessible to participants during interface usage. 
Next, participants used either the \precision{}, \rank{}, or \paint{} interface to convey scene insight in the form of possible object locations and their likelihoods to the robot.
Finally, participants were administered questionnaires and were optionally able to provide more details about their experience within a free-response text box.
After the completion of all three trials, participants filled out demographic information.

Note that for each trial, we generated a fresh ground truth distribution.
To generate ground truth distributions, waypoints in the office environment (see Figure \ref{fig:map}) were randomly sampled from a set of candidate waypoints.
The likelihood of an object being in a particular ground truth location was also randomly sampled.

\subsection{Measures}
We measure usability subjectively through the system usability (SUS) scale (10 items on a 5-point likert scale) \cite{brooke1996sus}, a common usability questionnaire that produces a usability score between 0 and 100.
We measure cognitive load subjectively based on \textit{mental demand}, \textit{hurriedness}, and \textit{difficulty} of the task.
Each factor of cognitive load is measured using a single 7-point Likert scale item drawn from the NASA Task Load Index (TLX) \cite{hart1988development}: \textit{``How mentally demanding was the task?''} (mental demand); \textit{``How hurried or rushed was the pace of the task?''} (hurriedness); and \textit{``How hard did you have to work to accomplish your level of performance?''} (difficulty). 
Additionally, our demographic questionnaire asked participants to rate their agreement to the statements \textit{``I am familiar with robots''} and \textit{``I have experience with computer programming''} on a 7-point Likert scale.

Our objective measures include \textit{efficiency} and \textit{accuracy}.
Efficiency is measured by the duration that participants spent using each interface.
Accuracy is measured along two dimensions.
First, \textit{distribution} accuracy compares the probability distributions produced by the \precision{} and \paint{} interfaces to their corresponding ground truths.
To calculate distribution accuracy, we vectorize the probabilities of both user-specified and ground-truth distributions and return the cosine similarity.
\Rank{} does not produce a distribution and cannot be included in this measure.

Second, \textit{rank discrepancy} compares all three interfaces.
Rank discrepancy converts the \precision{} and \paint{} distributions to ordinal rankings in which the highest-probability waypoint is ranked first.
We compare each user-created ranking to its corresponding ground-truth ranking via the Damerau-Levenshtein distance.
If a user's ranking encompasses all ground-truth locations, we consider only the shortest prefix of their ranking that contains all ground truths.

\subsection{Analysis}

We perform repeated-measures analyses of variance (ANOVA) for our usability (SUS) measure.
For the ANOVA, we test the assumption of sphericity using Mauchley's Test of Sphericity, and if violated, we apply a Greenhouse-Geisser correction.
Post hoc comparisons for usability are performed via the pairwise two-tailed t-test with Bonferroni correction.
In applying the Bonferroni correction, we multiply p-values by the number of comparisons, which in our case is $3$, and report the result.
We perform non-parametric tests for our single-item (mental demand, hurriedness, and difficulty), efficiency, accuracy, and performance measures.
The Friedman test is our non-parametric alternative to the repeated-measures ANOVA, and the two-tailed Wilcoxon signed-rank test is our non-parametric alternative to the pairwise t-test.
Non-parametric post hoc tests are similarly adjusted via the Bonferroni correction.
In all of our analyses, a p-value of $0.05$ is our threshold for significance, though we additionally report marginal results with p-values up to $0.1$.

\subsection{Results}

\begin{figure}[!b]
    \centering
    \includegraphics[width=\columnwidth]{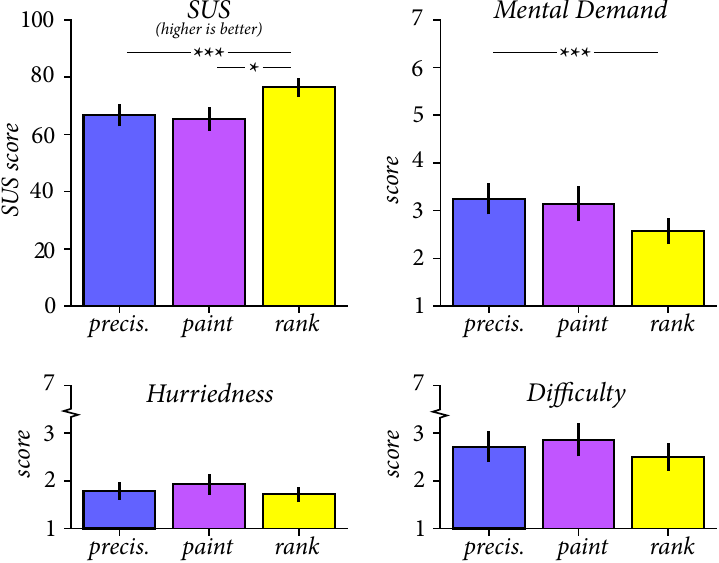}
    \caption{The results for our subjective measures. Error bars represent standard error of the mean. Lower values are better for all measures except for SUS. *p < 0.1, **p < 0.05, ***p < 0.01}
    \label{fig:subj}
\end{figure}

\subsubsection{Participants}

We recruited $28$ naïve participants ($18$ Male, $10$ Female) from within the U.S. Naval Research Laboratory for our evaluation. The average age of participants was $40.3$ (SD=$11.3$). On a scale of 1 to 7, the average robotics familiarity among participants was $4.46$ (SD=$1.77$), while the average programming experience among participants was $4.82$ (SD=$2.21$).

\subsubsection{User Experience}

Figure \ref{fig:subj} (top left) depicts our usability results. In applying a repeated-measures ANOVA to analyze usability, the assumption of sphericity was violated according to Mauchley's Test of Sphericity, $\chi^2(2)=10.6$, p < 0.01.
We therefore applied a Greenhouse-Geisser correction and detected a significant effect of interface on usability, F(2, 54) = 3.97, p = 0.037.
Post hoc comparisons using the paired t-test with Bonferroni correction indicate a significant difference between the \rank{} ($M=76.4$, $SD=17.1$) and \precision{} ($M=66.6$, $SD=20.3$) interfaces, t(27) = 3.33, p < 0.01, and a marginal difference between the \rank{} and \paint{} ($M=65.4$, $SD=22.1$) interfaces, t(27) = -2.50, p = 0.056.

Figure \ref{fig:subj} illustrates the results for mental demand (top right), hurriedness (bottom left), and difficulty (bottom right). 
We detected a significant effect of interface on mental demand, $\chi^2(2)=8.94$, p = 0.011.
Post hoc comparisons for mental demand indicate a significant difference between the \rank{} ($M=2.57$, $SD=1.45$) and \precision{} ($M=3.25$, $SD=1.73$) interfaces, Z = 2.72, p < 0.01.
No significance was detected in hurriedness or difficulty.

\subsubsection{Efficiency}

\begin{figure*}[!t]
    \centering
    \includegraphics[width=\textwidth]{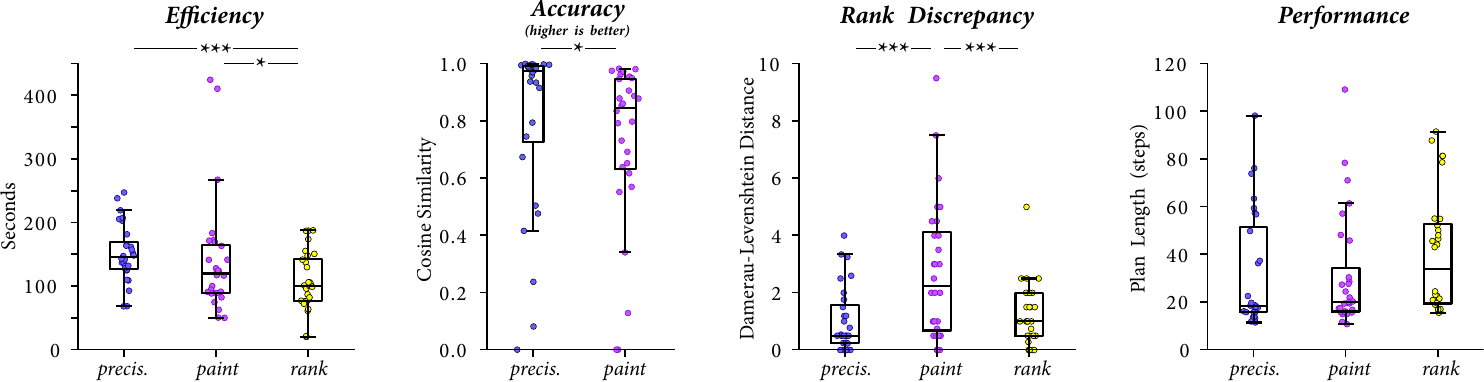}
    \caption{Box plots depicting our objective results. Lower values are better except for accuracy. *p < 0.1, ** p < 0.05, *** p < 0.01}
    \label{fig:obj}
\end{figure*}

Figure \ref{fig:obj} (left) depicts the results for efficiency.
We detected that the type of interface significantly affects efficiency, $\chi^2(2)=16.3$, p < 0.001.
Post hoc comparisons indicate a significant difference between the \rank{} ($M=108.8$, $SD=45.7$) and \precision{} ($M=151.5$, $SD=45.3$) interfaces, Z = 3.94, p < 0.001.
A marginal pairwise difference was additionally detected between \rank{} and \paint{} ($M=140.7$, $SD=91.0$), Z = 1.73, p = 0.084.

\subsubsection{Accuracy}

Figure \ref{fig:obj} (center left) depicts our distribution accuracy results.
We detected a marginal difference between \precision{} ($M=0.80$, $SD=0.30$) and \paint{} ($M=0.73$, $SD=0.29$) interfaces when comparing scene insight to ground truth, Z = 1.84, p = 0.066.
We also detected a significant effect on rank discrepancy (Figure \ref{fig:obj}, center right), $\chi^2(2)=11.9$, p < 0.01.
Post hoc comparisons indicate significance between \paint{} ($M=2.74$, $SD=2.41$) and \precision{} ($M=1.06$, $SD=1.14$), Z = 3.03, p < 0.01, and between \paint{} and \rank{} ($M=1.23$, $SD=1.10$), Z = 3.15, p < 0.01.

\subsection{Planning Case Study}

We now demonstrate how a robot can leverage scene insight in order to complete a task.
Consider the mobile robot from our evaluation procedure in section \S\ref{sec:procedure}.
The robot is given the goal that \textit{the umbrella must be placed in the bag}, but the locations of the umbrella and the bag are unknown to the robot.
End users have provided the robot with hints about the locations of these objects in the form of probability distributions (\ie{} from the \precision{} or \paint{} interfaces) or location rankings (\ie{} from the \rank{} interface).

This case study makes several assumptions about the robot's capabilities and limitations.
First, the robot navigates from waypoint to waypoint in the topographic mesh depicted in Figure \ref{fig:map} and can only observe objects that it is co-located with at the same waypoint.
Additionally, the robot has perfect perception within a waypoint; that is, if co-located with an object, the robot is assumed to be able to recognize it. 
Lastly, the starting position of the robot is assumed to be in the upper-right corner of the workspace room.

For this case study, our choice of planner is \textit{FF-Replan} with single-outcome determinization \cite{yoon2007ff}.
Although not an optimal probabilistic planner, FF-Replan is regarded as a high-performing baseline for the problems that \tool{} is intended to handle \cite{little2007probabilistic}.
Crucially, FF-Replan requires only a small amount of computational overhead, making it a realistic choice for mobile-robot applications where computational resources are limited. By contrast, optimal planners such as \textit{POMCP} are much more computationally intensive \cite{silver2010monte}.

In our implementation of FF-Replan, the robot first determinizes the planning domain using the most-likely object locations.
This determinization is the robot's initial \textit{belief}.
Note that this determinization process works for both probability distributions and rankings.
Next, the robot produces a plan, and executes the plan until making an observation that contradicts its belief (\ie{} a task-critical object not being where the robot expects it to be).
The robot then pauses execution, updates its belief by selecting the next most-likely location for that object, and replans.
If the robot exhausts its scene insight without finding both objects, it searches the remaining unvisited waypoints in a nearest-neighbor fashion.

We ran 50 planning simulations for each instance of scene insight provided by participants.
In each simulation, we sampled a ground-truth state from the ground-truth distribution paired with each instance. 
We then executed FF-Replan with the user-provided scene insight guiding plan creation and the ground truth state dictating the robot's observations.
The length of the resulting execution trace corresponds to the quality of the scene insight.
Shorter traces are better because the robot completes its task quicker.

We score user-provided scene insight by computing the average length of the 50 plans. The resulting scores are then averaged across each interface.
Shown in Figure \ref{fig:obj} (right), the average score is 32.3 ($SD=24.3$) for the \precision{} interface, 31.3 ($SD=24.0$) for the \paint{} interface, and 40.3 ($SD=25.0$) for  the \rank{} interface.
A Friedman test indicates that interface has an effect on performance, $\chi^2(2)=6.0$, p = 0.05. However, no pairwise differences were detected. 

%% file: 5discussion.tex
\section{Discussion}

Each interface exhibits several advantages and disadvantages.
As expected, efficiency is highest with the \rank{} interface and lowest with the \precision{} interface, supporting hypothesis \hfou{}. 
However, while \precision{} expectedly has significantly lower usability and higher cognitive load than \rank{}, the \paint{} interface was not found to be the most usable or the least cognitively demanding.
Thus, hypotheses \htwo{} and \hthr{} are unsupported.
Interestingly, the \paint{} and \precision{} interfaces differed only marginally in terms of accuracy, and \paint{} resulted in the highest rank discrepancy score.
Hypothesis \hfiv{} is thus unsupported.
Overall, there exist significant and marginal effects of interface choice on usability, cognitive load, efficiency, and accuracy, rendering hypothesis \hone{} partially supported. 
We discuss the implications of these results below.

\subsection{User Experience and Efficiency}

Our results indicate that the \rank{} interface outperforms \precision{} for usability and mental demand.
Furthermore, \rank{} enables users to convey scene insight to the robot more efficiently than with \precision{}.
On average, the \paint{} interface produced results that closely align with its counterparts, but we cannot conclude whether it performs better or worse.
The primary implication from these findings is that \textit{\textbf{from the user's perspective, ranking is a better option than expressing precise probability distributions due to affording faster and easier expression of scene insight.}}

\subsection{Implications for Robot Planning}

In terms of accuracy and rank discrepancy, neither the \precision{} nor \rank{} interfaces significantly outperformed each other.
Had \precision{} outperformed \rank{}, we would have concluded that choosing the right interface constitutes a tradeoff between user friendliness and accuracy; in reality, we lack evidence for such a tradeoff.
Our case study similarly shows no significant difference between \precision{} and \rank{}.
Thus, there is no strong evidence that such a tradeoff exists for this task.
The second implication of our work is that \textit{\textbf{while the \precision{} interface allows robot end users to specify exact probability distributions, we have little evidence to suggest that this precision matters in practice for these kinds of tasks.}} 
More work is needed to uncover the conditions under which the \precision{} interface produces better scene insight for other tasks.

\Paint{}, by contrast, performed significantly worse than the other interfaces in rank discrepancy and marginally worse than \precision{} in accuracy.
On further inspection of the heat maps created by participants, this difference is largely expected based on \paint{}'s mechanics.
With \paint{}, the likelihood of an object existing at a waypoint is calculated on a pixel-by-pixel basis, rather than by its proximity to individual user-specified points as with \precision{} and \rank{}.
Thus, \paint{} results in a greater number of waypoints with non-zero probabilities, leading to a higher rank discrepancy and marginally lower accuracy when compared to ground truth.
Effectively, \paint{} is sloppier than the other interfaces.

Interestingly, the case study's performance deviates from rank discrepancy and accuracy.
Within the sample that we collected, \paint{} performs nearly the same as \precision{} and better than \rank{}.
Given \paint{}'s significantly worse rank discrepancy and marginally worse accuracy, one would expect \paint{} to perform worse in the case study.
To explain this deviation, we hypothesize that the robot actually benefits from the \paint{} interface's sloppiness at runtime.
That is, if provided imprecise scene insight, the robot searches for an object in the general vicinity of a location rather than at a single waypoint, ultimately making search more resilient to user error.
We note that this result likely depends on the assumptions of our case study.
Tasks that require high precision and have high cost of errors (\eg{} fine-grained manipulation tasks) may suffer from \paint{}'s sloppiness.
Conversely, any benefit to \paint{}'s sloppiness may be diminished in environments with more physical space between waypoints.
Our final implication is therefore that \textit{\textbf{while painting heat maps is demonstrably less accurate than other approaches for expressing scene insight, this lack of accuracy \textit{may} not matter in some contexts.}}

\subsection{Limitations and Future Work}

Our work has several limitations.
First, our evaluation was conducted via a web interface rather than in-person. 
Although we controlled participants' devices to be either a desktop or laptop, the online nature of our study still may have introduced noise in our data.
It also limited our ability to test the \tool{} on alternate interface paradigms, such as handheld mobile devices.
In the future, we plan to investigate the performance of different uncertainty expression paradigms both in person and on alternate hardware platforms.

Several interface design choices may have also affected our results.
For example, a few participants expressed the \precision{} interface's normalization procedure (\S\ref{sec:design_precision}) to be confusing.
Some participants suggested that being unable to control the precision in the \paint{} interface (\eg{} by changing the brush size) affected their experience.
Future work must thereby further validate our results with alternate interface design choices.

The novice composition of our study population comprises a further limitation.
Although our findings primarily pertain to novice use, we believe that the tutorial about basic controls (\S\ref{sec:procedure}), the simplicity of each interface, and the similarity between each interface mitigates novelty effects in our data.
Still, further testing is required with more experienced users in order to strengthen our results.

Lastly, while our case study offers a glimpse of whether robot performance may be affected by interface choice, it only tests performance under a specific set of assumptions, and it does not test on a physical robot platform or using an optimal probabilistic planner.
Our immediate future work will test each approach for extracting scene insight on physical robot platforms equipped with optimal probabilistic task planners.
Ultimately, expanding our performance tests informs future improvements on state-of-the-art robotics and planning techniques.